# A Solution-Processed Ti/Pt/$NiO_x$/Ag RRAM Device with Multiple Forming-Free Non-Volatile Intermediate States

Abhrajit Roy[1], Chopali Chanchal Sahu[1], Rudraksh Sangore[1], Raman Hissariya[1], Poulomi Chakrabarty[2], Harsh Roniyar[1], Merlyne De Souza[3], and Debanjan Bhowmik[1,4]

[1]*Department of Electrical Engineering, Indian Institute of Technology Bombay, Mumbai, Maharashtra 400076, India*

[2]*Department of Computer Science and Engineering, Shiv Nadar University Chennai, Kalavakkam, Tamil Nadu 603110, India*

[3]*School of Electrical and Electronic Engineering, The University of Sheffield, Western Bank, Sheffield, S10 2TN, United Kingdom*

[4]*Centre for Semiconductor Technologies (SemiX), Indian Institute of Technology Bombay, Mumbai, Maharashtra 400076, India*

## Abstract

Artificial neural networks (ANN) implemented on crossbar arrays of non-volatile synaptic devices are known to consume much less energy compared to that implemented on conventional computing units. Solution-processed resistive random-access memory (RRAM) devices based on transition-metal oxides are suitable as such synaptic devices because of their cheap and easy fabrication method, stability, and suitable memristive properties. Here, in NiO-based RRAM devices with Ti/Pt as the bottom electrode and Ag as the top electrode (NiO layer deposited through solution processing), we experimentally report forming-free binary switching between a high resistance state (HRS) and a low resistance state (LRS) as well as forming-free pulse-driven modulation (long-term potentiation (LTP), long-term depression (LTD)) through 30 intermediate synaptic states between HRS and LRS (~ 5-bit synaptic weight resolution). Many of these states are measured to be stable at least up to $10^3$ seconds confirming their non-volatility and their utility in storage of synaptic weights for ANN crossbar arrays. On the contrary, most prior experimental reports (including reports both on solution-processing-based and sputtering-based RRAM devices) require forming to exhibit non-volatility. Avoiding forming reduces variability, eliminates the need for high voltage pulses, and makes the crossbar array design simpler. Further, we confirm the presence of oxygen vacancies in our materials stack using X-ray photo-electron spectroscopy (XPS). Finally, we show successful data classification on MNIST and CIFAR-10 data sets using a multi-layer perceptron (MLP) model and a convolution neural network (CNN) model respectively, where the synaptic weight updates of these ANN models are calibrated against the experimentally measured LTP-LTD characteristics of the NiO RRAM devices.

## 1. Introduction

Artificial neural network (ANN) models of the feedforward type, like multi-layer perceptron (MLP) and convolutional neural networks (CNN), have been found extremely useful for identifying objects inside images. However, implementing these ANN models on conventional digital silicon-technology-based computing systems like central processing unit (CPU) and graphics processing unit (GPU) involves relatively low speed and high energy consumption due to the inherent memory-computing separation in these systems, also known as von Neumann bottleneck. Compute-in-memory (CIM) or neuromorphic crossbar arrays of non-volatile memory (NVM) synaptic devices may come to the rescue because they intertwine memory and computing, and it has been shown experimentally that implementing such ANNs on these crossbar arrays is much faster and energy-efficient [1–4].

Among the various emerging NVM devices that have been considered for such crossbar arrays, resistive random-access memory (RRAM, also abbreviated as ReRAM) devices based on binary transition-metal oxides are very popular because of their simple two-terminal metal insulator-metal structure, high material stability, and suitable memory properties [5–8]. Nickel oxide (NiOx) is a popular binary transition-metal oxide in this regard and can be deposited through both the high-vacuum sputtering technique [9–12] or the solution processing technique [13–18] with the latter technique being much cheaper compared to the former.

But to the best of our knowledge, most prior reports on NiOx RRAM devices, whether the NiOx layer is deposited through sputtering or solution processing, show the requirement of an electro-forming step (where a voltage much higher than the switching voltage is applied to create the filamentary pathway in the oxide layer) to achieve non-volatile bipolar switching between a high resistance state (HRS) and a low resistance state (LRS) [9,10,13,18]. This is shown in Table 1. Intermediate non-volatile states between HRS and LRS, which can be used to store ANN weight values when used as NVM synapses in the crossbar array, have also been reported in these devices only after carrying out the electroforming step [9,10,15,16,18], as shown in Table 2. If intermediate states are measured through the application of pulses before carrying out forming, the states are found to be volatile because of the formation of weak filaments that disintegrate [13,14], leading to weight decay (inappropriate for ANN implementation on crossbar arrays).

However, carrying out electro-forming first to operate RRAM devices as NVM synapses in a crossbar array requires complicated peripheral circuitry to apply a large forming voltage and handle device-to-device variations related to the forming voltage [22,23]. Also, unselected devices in the crossbar array can be affected by the forming voltage due to sharing of the same word or bit line.

In that context, in this article, we experimentally report in solution-processed NiOx-based RRAM devices (with Ti/Pt as the bottom electrode (BE) and Ag as the top electrode (TE)) the occurrence of non-volatile states without the need for an electro-forming step. First, we demonstrate binary switching between non-volatile LRS and HRS states (memristive behaviour) in these devices across $10^3$ cycles with low cycle-to-cycle variation, which we have characterized here. LRS and HRS states are shown to be stable up to $10^4$ s. We have also demonstrated this switching across 15 devices, fabricated using the same technique, and reported very low device-to-device variation. From the binary switching characteristics of the device between HRS and LRS, through appropriate curve fitting, we identify different transport regimes: Ohmic, Schottky emission, and trap-controlled space-charge-limited conduction (T-SCLC). We have also confirmed the presence of oxygen vacancies in the material using X-ray photo-electron spectroscopy (XPS) and thus confirmed that filaments of oxygen vacancies do contribute to this memristive behaviour [24–27].

Furthermore, without the requirement of a forming step, we demonstrate pulse-driven modulation (long-term potentiation (LTP), long-term depression (LTD)) spanning 30 intermediate Table 1: Comparison of previously reported NiOx RRAM devices and the device in our work in terms of their current vs voltage characteristics, which show high resistance state (HRS) and low resistance state (LRS). Materials stack is listed for each device with bottom electrode (BE) first and top electrode (TE) last. Devices with oxide layer deposited by solution processing as well as devices with oxide layer deposited by sputtering are included in this comparison.

*Table 1: Comparison of previously reported NiO RRAM devices and the device in our work in terms of their current vs voltage characteristics, which show high resistance state (HRS) and low resistance state (LRS). Materials stack is listed for each device with bottom electrode (BE) first and top electrode (TE) last. Devices with oxide layer deposited by solution processing as well as devices with oxide layer deposited by sputtering are included in this comparison.*

| **Materials Stack and Reference** | **HRS** | **LRS** | **Set voltage** | **Reset voltage** | **Min. retention time** | **Electro-forming step required?** |
|---|---|---|---|---|---|---|
| Ti (20 nm) / Pt (80 nm)/ NiO (50 | ≈100 $\mu$S | ≈ 5 mS | ≈ 0.5 V | ≈ -0.7 V | $10^4$ s | Yes (≈ 1.25 V) |
| nm)/ Ag (160 nm): Li *et al.* [13] | | | | | | |
| Ti (20 nm) / Pt (80 nm)/ NiO (50 nm)/ Ag: Chu *et al.* [18] | ≈10 nS | ≈ 5 mS | ≈ 0.35 V | ≈ -1.5 V | $10^4$ s | Yes (≈ 1.5 V) |
| Au/ NiO/ Au: Hadiyal *et al.* [14] | Not clearly mentioned | Not clearly mentioned | Not clearly mentioned | Not clearly mentioned | Not mentioned | Not clearly mentioned |
| ITO/ NiO (23 nm)/ Ag: Kanan *et al.* [16] | ~10 $\mu$S | ~10 mS | ~5 V | ~-2.2 V | $10^4$ s | Yes |
| Nb/ NiO (50 nm)/ Nb: Ahn *et al.* [19] | ~60 mS | ~40 mS | – | – | – | Yes (4.1 V) |
| ITO/ NiO (23 nm)/ Ag: Park *et al.* [20] | ~2 mS | ~28 mS | ~1.8 V | ~-1.2 V | 1200 s | Yes |
| Pt/$NiO_y$(20 nm)/$NiO_x$(20 nm)/Cu: Sarkar *et al.* [9] | ~0.2–1 $\mu$S | (1 mA) ~0.5–1 mS (3 mA) ~1.7–3.3 mS (5 mA) ~5–10 mS | ~2.5 V | ~1.5 V | $10^4$ s | Yes (~4.4 V) |
| Pt/ NiO (50 nm)/ Cu: Ma *et al.* [10] | ~50 mS | ~10 nS | ~0.4 V, ~0.9 V | -1.5, -0.9, -0.6, -0.4 V | over $10^4$ s | Not clearly mentioned |
| W (3 nm) / Pt (4 nm) / Co (0.8 nm)/ NiO (30 nm): Hsin *et al.* [21] | ~70 nS | ~1.0 mS | 2.7 V | ~2 V | $10^4$ s | Not clearly mentioned |
| **Ti (20 nm) / Pt (80 nm)/ NiOx (17 nm)/ Ag (100 nm): This work** | ≈ 2.5 mS | 70 mS | ≈ 0.5 V | ≈ -0.5 V | $10^4$ s | Forming free |

states (≈ 5-bit synaptic weight resolution). Many of these states are measured to be stable at least up to $10^3$ seconds, thereby confirming their non-volatility and their utility in storage of synaptic weights for the aforementioned crossbar arrays. We use identical positive pulses (0.3 V magnitude) for LTP and identical negative pulses (0.3 V magnitude) for LTD (Table 2). This identical pulse scheme makes peripheral circuit design for programming the synapses in the crossbar array simple.

On the contrary, to the best of our knowledge, very few non-volatile intermediate states (only about 2-3 states) with high retention time have been reported in both solution-processed and sputtered NiOx devices even after carrying out the electroforming step first (Table 2) [9,10,15,16,18]. Also, in these reports, the intermediate states have been created by applying different set or reset voltages or different compliance currents and thereby obtaining minor hysteresis loops [10, 16, 18]. Programming devices in this manner requires very complicated peripheral circuitry, which isn't the case in our work since we use the identical pulse scheme (as mentioned above).

*Table 2: Comparison of previously reported NiOx RRAM devices and the device in our work in terms of intermediate states exhibited between HRS and LRS. Materials stack is listed for each device with BE first and TE last. Devices with oxide layer deposited by solution processing as well as sputtering are included.*

| Materials Stack and Reference | When are intermediate states created? | How are intermediate states are created? | Number of intermediate states | Conductance range of intermediate states | Stability of intermediate states |
|---|---|---|---|---|---|
| Ti (20 nm) / Pt (80 nm)/ NiO (50 nm)/ Ag: Chu *et al.* [18] | After forming | Adjusting reset voltages: -0.5 V, -1.2 V, -2 V | 4 | Lowest: ~0.5 $\mu$S, Next: ~50 $\mu$S, Next: ~0.5 mS, Highest: ~2 mS | Non-volatile (retention: $10^4$ s or more) |
| Pt/NiO$_y$(20 nm)/NiO$_x$(20 nm)/Cu: Sarkar *et al.* [9] | After forming | Adjusting compliance current: 1 mA, 3 mA, 5 mA | 3 | Compliance 1 mA: 0.5–1 mS, 3 mA: 1.7–3.3 mS, 5 mA: 5–10 mS | Non-volatile (retention: $10^4$ s or more) |
| Pt/ NiO (50 nm)/ Cu: Ma *et al.* [10] | Not clearly mentioned | Adjusting reset voltages: -1.5 V, -0.9 V, -0.6 V, -0.4 V | 4 | -1.5 V: 1–10 nS, -0.9 V: 0.3–1 $\mu$S, -0.6 V: 0.1–0.2 mS, -0.4 V: 1.7–5 mS | Non-volatile (retention: $10^4$ s or more) |
| ITO/ NiO (70 nm)/ Al (120 nm): Kanan *et al.* [16] | Not clearly mentioned | Adjusting reset voltages: 1 V, 0.8 V, 0.6 V, 0.5 V | 4 | 0.5V: ~14 $\mu$S 0.6V: ~0.11 mS, 0.8V: ~1.3 mS, 1.0V: ~10 mS | Non-volatile ($10^4$ s) |
| Ti (10 nm)/ Pt (70 nm)/ NiOx (96 nm)/ Pd (200 nm): Mondal *et al.* [12] | Not clearly mentioned | Voltage pulses (500 ms duration) | 100 | ±5 mV pulses: 0.21–0.24 mS, ±10 mV pulses: 0.20–0.26 mS, ±15 mV pulses: 0.19– | Volatile (decays in ≈ 10 s) |
| | | | | 0.28 mS. ±20 mV pulses: 0.18–0.30 mS | |
| Ti (20 nm) / Pt (80 nm)/ NiO (50 nm)/ Ag (160 nm): Li *et al.* [13] | Before forming | ±0.7 V, 2 ms pulses | 50 | ≈32 $\mu$S–35 $\mu$S | Volatile (decays in ≈400 s) |
| ITO/ NiO (23 nm)/ Ag: Park *et al.* [20] | Before forming | Voltage pulses | 80 | ±0.8 V pulses (10 ms): 0.5–7.5 $\mu$S, ± 1.0 V pulses (10 ms): 0.7–9.0 $\mu$S, 1 V pulses (20 ms): 1.3–7.5 $\mu$S | Volatile (decays in ≈250 s) |
| Au/ NiO/ Au: Hadiyal *et al.* [14] | Before forming | ±0.5 V, 1 V, 1.5 V, 2 V (75 ms) pulses | 25 | 0.5 V pulses: 10–12 mS, 1 V pulses: 21–24 mS, 1.5 V pulses: 34–39 mS, 2 V pulses: 50–57 mS | Volatile (decays in ≈100 s) |
| **Ti (20 nm) / Pt (80 nm)/ NiOx (17 nm)/ Ag (100 nm): This work** | Forming free | ±0.3 V pulses of 200 ns duration each | 30 | ≈ 3 mS to 30 mS range | Non-volatile (retention: $10^3$ s or more) |

Finally, we show successful data classification on the MNIST data set comprising images of handwritten digits using a multi-layer perceptron (MLP) model and on the CIFAR-10 data set comprising images of different animals and transport vehicles using a convolution neural network (CNN) model [28–30]. We have simulated the MLP and the CNN in the on-chip inference mode [1–4], with the synaptic weights of these ANN models calibrated against our experimentally measured LTP-LTD characteristics of the NiOx RRAM devices during inference (forward computation of the trained ANN).

## 2 Materials and Methods

### 2.1 Thin Film Deposition and Characterization

The Si substrate is first cleaned and thermally oxidized to form a 200 nm-thick $SiO_2$ insulating layer. A 20 nm thick Ti layer followed by an 80 nm thick Pt layer is deposited through high vacuum DC sputtering; this bilayer acts as the bottom electrode in the final device.

To deposit the active layer (NiOx), solution processing technique is carried out as described next. To prepare the NiOx precursor, 290.8 mg of $Ni(NO_3)_2$,$6H_2O$ is dissolved in 5 mL of deionized water and stirred (using a magnetic stirrer) at 85 C for 1 hour ensuring complete dissolution of the salt. Once a clear solution is obtained, 100 $\mu$L of $NH_4OH$ is added drop-wise to the $Ni(NO_3)_2$ solution amidst continuous stirring to gradually increase the pH and trigger the nucleation of $Ni(OH)_2$. The solution is continued to be stirred at room temperature for 20 hours, allowing uniform precipitation of $Ni(OH)_2$. The obtained precipitate is collected and repeatedly washed through centrifugation to remove excess ions and residual nitrates, resulting in a clean green $Ni(OH)_2$ gel.

This NiOx precursor is then deposited on the Si substrate through spin coating and baking. To ensure uniform wetting of the NiOx precursor during coating, the oxidised Si substrate surface is treated with oxygen plasma to make it hydrophilic. Each spin-coating step is followed by a 10-minute bake at 120°C in ambient air to remove residual solvent and improve film uniformity. The oxygen plasma ashing, spin-coating, and bake cycle have been repeated five times to achieve the desired film thickness. A post-deposition annealing step is then carried out at 350°C for 60 minutes in air to induce film densification, enhance crystalline nature, and regulate oxygen vacancy distribution within the oxide layer. This entire process flow to deposit the Ti/Pt/NiOx thin film has been depicted through a schematic in Figure S1 of Supporting Information (Section S1). The final thin film stack is shown in Figure 1(a).

Subsequently, on the deposited thin film, high-resolution X-ray diffractometry (HRXRD) is carried out to understand the growth quality and X-ray reflectivity (XRR) is carried out to measure the thickness values of the deposited layers. Rigaku Smart-Lab XRD system is used for both XRD and XRR. X-ray photoelectron spectroscopy (XPS) is performed using PHI 5000 Versa-Probe II system on the same thin film to confirm the presence of oxygen vacancies. Further details on settings during these measurements are presented in Section S1 of Supporting Information. Thin film characterization results based on these techniques are presented in a later section of the main text and in Figure 1.

### 2.2 Device Fabrication and Characterization

Next, to fabricate two-terminal RRAM devices from the deposited thin film, 100 nm thick Ag layer is deposited in the form of 50 $\mu$m diameter circles on the thin film. Photo-lithography and DC sputtering have been carried out for the purpose (more details provided in Section S1 of Supporting Information). The final metal-insulator-metal (MIM) device structure is shown in Figure 2(a). This device fabrication process flow is shown through a schematic in Figure S2 of Supporting Information.

Details of the top electrode and bottom electrode in our device stack are further listed and compared with that used in other related reports on solution-processed NiOx devices in Table ST1 of Supporting Information. Details of our solution processing and spin coating steps to

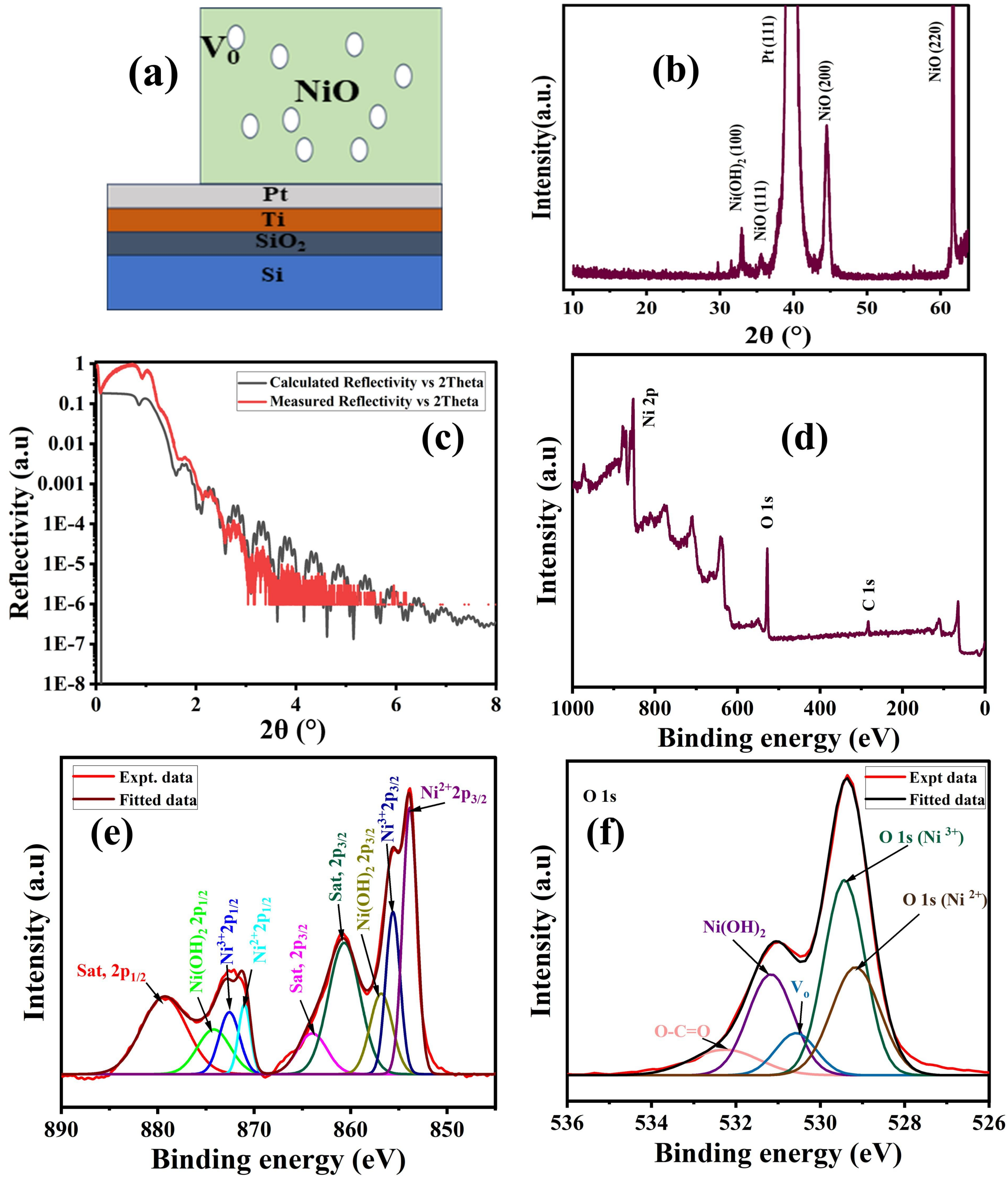

*Figure 1: (a) Schematic of the Si/SiO₂/Ti/Pt/NiOx thin film stack, not drawn to scale: thickness values of the layers mentioned in the text (white circles in the NiOx layer represent oxygen vacancies). (b), (c) Structural characterization of the same thin film using XRD and*

*XRR respectively. (d) XPS survey scan of the NiOx film showing Ni, O, and C signals. (e) High resolution Ni 2p XPS spectrum with $Ni^{2+}$, $Ni^{3+}$, $Ni(OH)_2$, and satellite peaks. (f) O 1s XPS spectrum showing lattice oxygen, $V_o$ (oxygen vacancy), $Ni(OH)_2$, and O–C=O components.*

deposit the oxide layer are further listed and compared with that followed in related reports on solution processed NiOx devices in Table ST2 of Supporting Information. Details of the materials stack and fabrication process for sputtered NiOx devices reported in literature is provided in Table ST3.

Electrical characterization of the devices to report their binary switching (Figure 2) and intermediate states (LTP-LTD) (Figure 4) is carried out using Keysight B1500A semiconductor parameter analyzer, equipped with a waveform generator/fast measurement unit (WGFMU).

### 2.3 Simulation of On-Chip ANN Inference

MLP and CNN models are designed and trained using the Keras library of Python programming language. Non-linearity factors for LTP and LTD of the device are extracted using the NeuroSim simulator [31,32]. On-chip inference of MLP and CNN is simulated in Python after mapping the weights of the trained MLP and CNN into the conductance values of the synaptic devices that are based on these obtained non-linearity factors (more details and results presented in a later section of the text and in Figure 5).

## 3 Results and Discussion

### 3.1 Materials Characterization: Confirmation of Presence of Oxygen Vacancies

As mentioned earlier, on the Ti/Pt/NiOx thin film (Figure 1(a)), before deposition of Ag top electrodes, XRD, XRR, and XPS measurements have been carried out to characterize the material stack and confirm presence of oxygen vacancies. Details of the instruments used and measurement settings for these measurements have been provided in Section S1 of Supporting Information. We present the measurement results below.

XRD diffraction pattern shown in Figure 1(b), confirms a predominantly face centred cubic (fcc) NiOx phase with broadened reflections centered at $2\theta = 37.23^{\circ}$, $43.28^{\circ}$, and $62.84^{\circ}$, corresponding to the (111), (200), and (220) planes respectively of cubic NiOx, respectively which are in excellent agreement with the standard diffraction data of fcc NiOx (JCPDS card no. 78-0643, space group $Fm\bar{3}m$). The significant attenuation and broadening of these reflections, when compared with bulk NiOx, indicate a nanocrystalline film morphology and strain effects induced by the underlying Pt(111) bottom electrode [33–35]. A distinct yet weak feature observed at approximately $2\theta \approx 33.1^{\circ}$ is attributed to the (100) plane of $\beta$-$Ni(OH)_2$. Such a reflection has frequently been reported in solution-processed and low-temperature-annealed NiOx films, signifying partial surface hydroxylation or incomplete conversion of the hydroxide precursor to NiOx [36,37]. The coexistence of the NiOx and $Ni(OH)_2$ signatures suggests that the surface of the oxide layer contains a thin hydroxide shell, likely formed due to post-deposition exposure to ambient moisture or residual precursor groups [36,37].

X-ray reflectivity (XRR) measurement (red plot), coupled with calculation/ fitting (black plot) performed using the Global Fit software [38], is shown in Figure 1(c). Fitting using thickness value of ≈ 17 nm (black plot) for the NiOx layer matches with experimental measurement. Density of the layer is considered to be 5.14 g/cm$^3$ and surface roughness ≈ 0.1 nm.

XPS spectrum of the same thin film is shown in Figure 1 (d), (e), (f). Dominant signals corresponding to the Ni 2p, O 1s, and C 1s core levels are visible in the rough scan of Figure 1 (d) which spans a wide range of binding energy. This is expected for a NiOx layer exposed to ambient conditions [39]. We next carry out finer or higher resolution XPS measurement around these Ni 2p, O 1s, and C 1s peaks and match our observations with previous reports [39–46]. Ni 2p and O 1s XPS scans are discussed here (Figure 1 (e), (f)), while the C 1s scan is discussed in Section S3 of Supporting Information.

Table 3: Percentage area contributions of the different peaks in the XPS spectrum for O 1 s in Figure 1(f)

| **Chemical Species** | **Peak Area** |
|---|---|
| Oxygen associated with $Ni^{2+}$ | 24.71% |
| Oxygen associated with $Ni^{3+}$ | 37.40% |
| Oxygen-vacancy component ($V_o$) | 8.02 % |
| O–C=O species | 8.23% |

As shown in the finer XPS spectrum measured around the previously observed Ni 2p peak (Figure 1(e)), a significant peak appears at around ~853.8 eV with a shoulder at ~855.5 eV, corresponding to the $Ni^{2+}$ ($2p_{3/2}$) and $Ni^{3+}$ ($2p_{3/2}$) states respectively [39,41]. In addition, a feature at ~856.7 eV is observed and can be attributed to the surface $Ni(OH)_2$, which commonly forms in solution-processed NiOx films. The higher-binding-energy region shows pronounced satellite peaks near ~860.6 and ~863.8 eV, confirming the presence of $Ni^{2+}$ species through well-known shake-up transitions. The $2p_{1/2}$ region demonstrates a clear $Ni^{2+}$ peak at ~ 870.9 eV, followed by the $Ni^{3+}$ component at ~872.6 eV, and a higher-binding-energy $Ni(OH)_2$ feature at ~874.1 eV, along with the associated satellite peak positioned at ~879.3 eV [41–46]. These observations collectively confirm that the deposited layer is composed of slightly non-stoichiometric NiOx containing $Ni^{2+}$, $Ni^{3+}$, and $Ni(OH)_2$ components.

In the finer XPS spectrum measured around the previously observed O 1s peak (Figure 1(f)), a predominant peak is observed at ~529.1 eV, corresponding to lattice oxygen ($O^{2-}$) bonded to $Ni^{2+}$, whereas the component at ~529.5 eV can be attributed to lattice oxygen associated with $Ni^{3+}$ species. The peak appearing at ~530.5 eV can be assigned to oxygen-vacancy–related states ($V_o$), in accordance with the report by Haunsbhavi *et al* [42]. This confirms the presence of oxygen vacancies, which play a significant role in digital and analog switching as discussed later. A higher-binding-energy feature at ~531.1 eV originates from $Ni(OH)_2$, which is typically formed due to surface hydroxylation in solution-processed NiOx. Additionally, the peak detected at ~532.2 eV corresponds to O–C=O contamination species [42,43,45,47].

By calculating the area under the curve, peak areas for the aforementioned chemical species in the O 1s XPS spectrum are listed in Table 3 [42]. As listed in Table 3, the oxygen-vacancy component ($V_o$) contributes to 8.02% of the net spectrum whereas lattice oxygen associated with $Ni^{2+}$ and $Ni^{3+}$ contribute to 24.71% and 37.40%. Thus, the ratio of oxygen vacancy component to oxygen associated with $Ni^{2+}$ is 0.325. The ratio of oxygen vacancy component to oxygen associated with $Ni^{3+}$ is 0.215. These numbers show the extent of oxygen vacancies in the deposited thin film stack. However, while our XPS spectrum measurement of the thin film stack (before top electrode deposition) discussed before (Figure 1 (d), (e), (f)) confirms the presence of oxygen vacancies, given that the deposited top electrode after that measurement is made of Ag, role of $Ag^+$ ions (conductive bridge) in filament formation in the final device (along with oxygen vacancies) and resultant memristive behaviour (as discussed next) cannot be ignored [13].

### 3.2 Device Characterization: Binary Switching between HRS and LRS

After depositing circular Ag top electrodes of diameter 50 $\mu$m on the Ti/Pt/NiOx thin film stack, device characterization is carried out. Bipolar resistive switching behavior is evident from the measured current (I)-voltage (V) characteristic (Figure 2(b)). Throughout the measurement the top electrode is biased at a DC voltage, which is varied slowly, while the bottom electrode is maintained at ground potential. Set transition (transition from HRS to LRS) is observed only under positive voltage sweep, while reset transition (transition from LRS to HRS) is observed only under negative voltage sweep as shown.

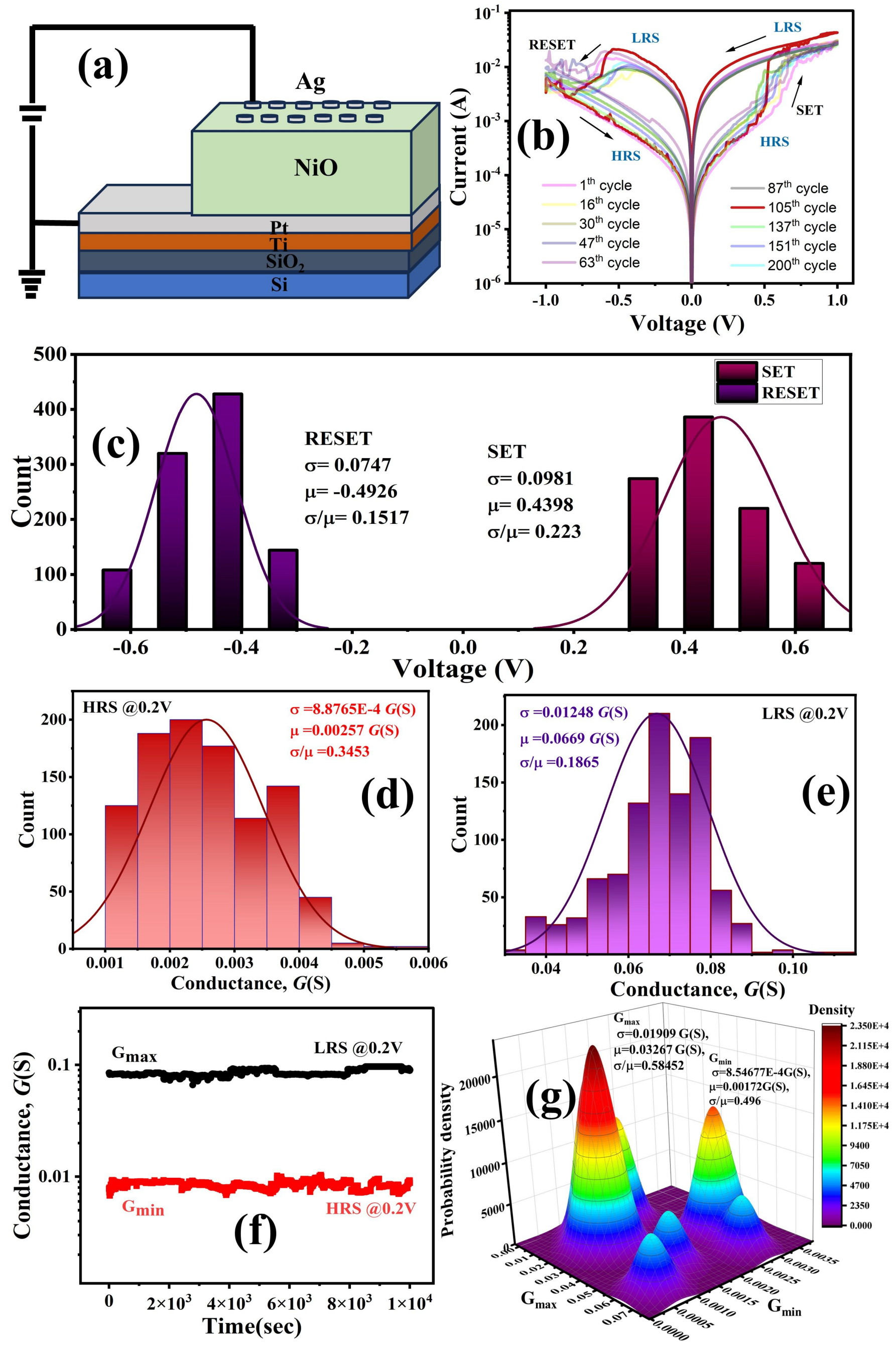

*Figure 2: (a) Schematic of the fabricated Si/SiO₂/Ti/Pt/NiOx/Ag device (b) Current vs voltage characteristic of the device for different cycles showing bipolar resistive switching: high*

*resistance state (HRS) to low resistance state (LRS) transition (set) at positive voltage, and LRS to HRS transition at negative voltage (reset). (c) Statistical distribution of set and reset voltages, extracted from 1000 switching cycles. (d), (e) Conductance distribution of HRS and LRS respectively, extracted from 1000 switching cycles. (f) Retention characteristics of HRS and LRS states, showing their stability (or non-volatility) up to $10^4$ s. (h) Probability density distribution of LRS state's conductance ($G_{max}$) and HRS state's conductance ($G_{min}$) measured from 15 different devices, thereby demonstrating device-to-device uniformity.*

In order to study cycle-to-cycle variations, we have repeated the above I-V characteristic measurement across 1000 cycles. I-V characteristics corresponding to different such cycles are plotted in Figure 2(b) and also in Figure S3 of Supporting Information (Section S2). Distribution of the set voltage (at which HRS to LRS transition happens) and reset voltage (at which LRS to HRS transition happens), measured across 1000 cycles, is plotted in Figure 2(c). Set and reset voltages are centered roughly around 0.5 V and -0.5 V respectively with very narrow dispersion. 0.5 V being a relatively low voltage magnitude, this shows that our device is suitable for low voltage and low energy operation. The limited variation in these switching voltage values indicates low cycle-to-cycle variability, which is essential for reliable device operation.

The distribution of conductance of HRS across 1000 cycles is shown in Figure 2(d) and that of conductance of LRS across 1000 cycles is shown in Figure 2(e). HRS conductance is centered around roughly 0.0025 S or 2.5 mS with a standard deviation of $8.88 \times 10^{-4}$ S, corresponding to a $\sigma/\mu$ value of about 0.3453. LRS conductance is centered around roughly 0.07 S or 70 mS (≈ 23 times higher than mean HRS) with standard deviation of 0.01248 S, yielding a $\sigma/\mu$ value of approximately 0.1865. The relatively low dispersion in LRS and moderate variation in HRS indicate stable conductive filament formation with some variability during rupture. Importantly, the two distributions remain well separated with negligible overlap and significant memory window, ensuring reliable distinction between the HRS and LRS states.

As shown in the retentivity characteristic of the device in Figure 2(f), HRS and LRS states are found to be stable at least up to $10^4$ s under continuous read bias. The maintenance of high ratio between conductance of LRS and HRS states ($G_{max}$/ $G_{min}$ ≈ 10X) and absence of noticeable drift in the conductance values confirm the non-volatile nature of this binary switching between HRS and LRS. As already mentioned in Section 1, we obtain these non-volatile HRS and LRS states without apriori carrying out an electro-forming step involving higher voltage application, in contrast to previous reports (Table 1) [9, 10, 13, 18]. We do not require forming in our device probably because of the relatively lower thickness of the oxide layer in our device (≈ 17 nm) compared to previously reported devices. Lower thickness of the oxide layer ensures that the filament formed, leading to LRS, does not disintegrate easily upon removal of voltage
(non-volatility).

Device-to-device uniformity in this binary switching is confirmed in through the probability density distribution plot of the LRS conductance ($G_{max}$) and HRS conductance values ($G_{min}$), as extracted from 15 different devices we have fabricated following the same process flow discussed above and plotted in Figure 2(g). The distribution of $G_{\max}$ is centered at 0.03267 S with a standard deviation of 0.01909 S, resulting in a $\sigma/\mu$ value of 0.58452, which corresponds to a variation of 58.45%. In contrast, the $G_{\min}$ distribution is centered at 0.00172 S with a standard deviation of $8.54677 \times 10^{-4}$ S, yielding a $\sigma/\mu$ value of 0.496, corresponding to a variation of 49.6%. Although both $G_{\max}$ and $G_{\min}$ show noticeable variation, the separation between the two distributions remains sufficiently large to maintain a clear distinction between the HRS and LRS states across different devices.

To understand the different conduction mechanisms at play during the above memristive switching, the current-voltage characteristics of the NiOx device have been further analyzed using double-logarithmic plots in Figure 3 (a), (b) (only magnitudes of applied voltages and measured currents are considered here). The conduction behavior can be divided into four different regimes during both the set and reset processes based on the extracted slopes, as discussed next. In regime I, the slope is close to 1, indicating Ohmic conduction, where the current shows an approximately linear relationship with the applied voltage [14,48–51]. In this low-voltage Ohmic conduction regime, the current is mainly governed by the thermally generated charge carriers present in the switching layer [48,49].

With an increase in voltage bias, the slope gradually increases to approximately 1.2; we demarcate this regime as regime II (Figure 3 (a), (b)). This slope does not correspond to typical Ohmic conduction regime. To further understand the conduction mechanism in this region,

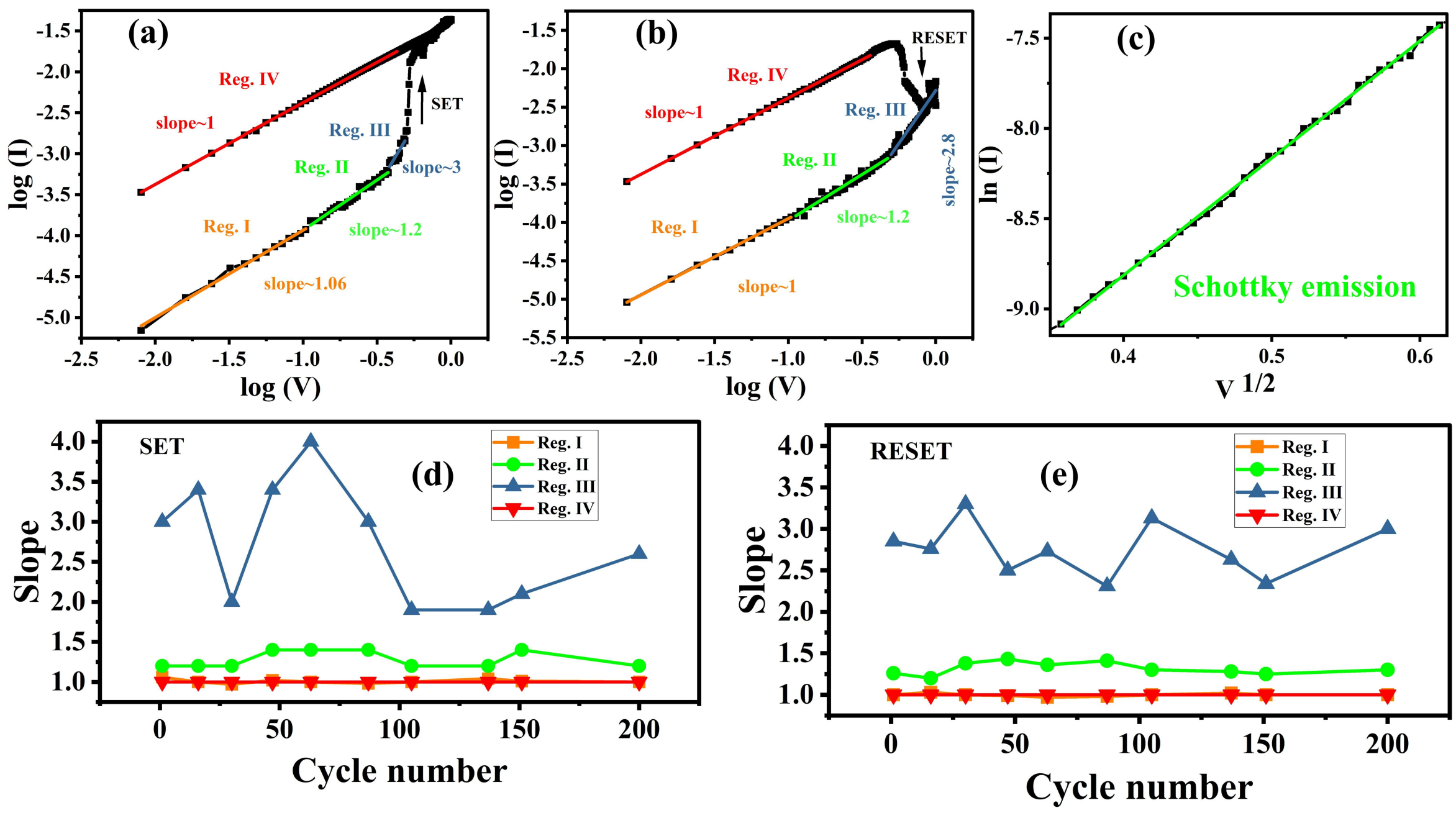

*Figure 3: (a), (b) Identification of different conduction regimes (Reg.) in the log(I)–log(V ) characteristics during set and reset operations respectively: regime I: Ohmic, regime II: Schottky emission, regime III: trap-controlled space-charge-limited conduction (T-SCLC),*

*regime IV: Ohmic. (c) For regime II, ln(I) varies linearly with $V^{1/2}$, further confirming Schottky emission behaviour, (d), (e) cycle-to-cycle variation of the extracted slopes during the aforementioned regimes for set and reset operations respectively.*

the current measurement data is replotted as ln($I$) versus $V^{1/2}$ in Figure 3(c). The linear dependence of ln($I$) on $V^{1/2}$, as suggested by the linear fitting with minimal error, indicates the contribution of Schottky emission to charge transport in this regime [14,49]. Carrier transport in this region is influenced by thermionic emission over the interfacial energy barrier, where the applied electric field facilitates carrier injection across the electrode/NiOx interface. Such Schottky-type transport associated with the metal/NiOx interface has also been reported in other reports on NiOx-based RRAM devices [13,14].

In the next regime, regime III, a much higher slope ( ≈ 3) is observed in the double logarithmic plot of Figure 3(a), (b), indicating strongly non-linear trap-controlled conduction. This behavior is consistent with trap-controlled space-charge-limited conduction (T-SCLC), where the injected carriers progressively occupy the available trap states, resulting in a rapid increase in current [49, 52–54]. In this regime, the device shifts from HRS to LRS during set (Figure 3(a)) and from LRS to HRS during reset (Figure 3(b)). The next regime obtained after the switching (regime IV) again shows a slope close to 1 for the double logarithmic plot (Figure 3(a), (b)), indicating Ohmic conduction once again.

The extracted slopes from the double logarithmic plot (Figure 3(a,(b)) for repeated set and reset cycles are shown in Figure 3 (d), (e) respectively. Regimes I, II, and IV exhibit relatively small variations in the corresponding slopes, while Regime III exhibits a larger variation in slope. This variation can be related to changes in the local defect configuration and conductive path during repeated switching cycles [13,54].

### 3.3 Device Characterization: Intermediate States, Pulse-Driven LTP-LTD

In Figure 4(a), conductance response of the same device in Figure 2(a) to 30 consecutive positive voltage pulses (+0.3 V, 200 ns duration) for LTP followed by 30 consecutive negative voltage pulses (-0.3 V, 200 ns duration) for LTD is shown. The LTP-LTD sequence is repeated over 8 cycles. The inset shows individual cycles while the main figure shows the mean across all cycles along with the variation range (maximum, minimum). The plot shows that cycle-to-cycle variation is minimal.

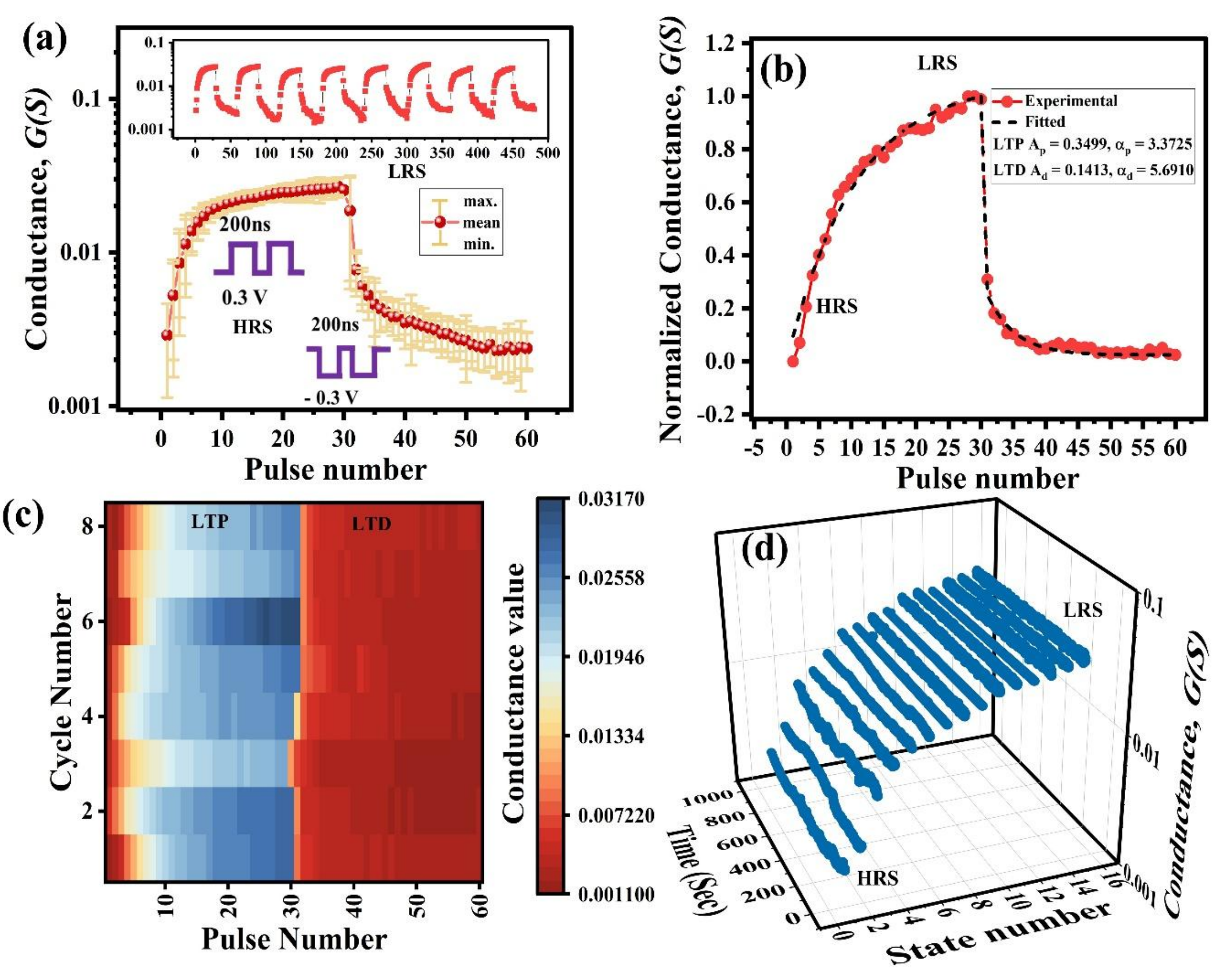

*Figure 4: (a) Conductance evolution of the NiOx RRAM device during eight consecutive LTPLTD cycles using identical pulses for LTP (+0.3 V, 200 ns) and identical pulses for LTD (-0.3 V, 200 ns). The inset shows the individual cycling behavior, while the main plot*

*represents the statistical distribution (minimum, mean, and maximum) across all cycles. (b) Normalized conductance of the 8th LTP/LTD cycle demonstrating synaptic weight update characteristics and fitting of non-linearity factors on them. (c) Heatmap representation of conductance evolution across eight cycles, illustrating pulse-dependent modulation and cycle-to-cycle variation during LTP and LTD processes. (d) Retention characteristics of 15 intermediate conductance states (selected from 30 states), confirming the non-volatile and stable multilevel switching behavior of the device.*

LTP-LTD for the 8th cycle of the inset of Figure 4(a) is plotted separately in Figure 4(b) after normalizing the conductance between 0 ($g_{min}$) and 1 ($g_{max}$). Following the method followed by the NeuroSim simulator [31,32], the LTP data is fitted using the following equation:

$$G(p_1) = \frac{g_{\max} - g_{\min}}{1 - e^{-1/A_p}} \left(1 - e^{-\frac{1}{A_p}\frac{p_1}{p_{\max}}}\right) + g_{\min}, \quad (1)$$

where $A_p$: non-linearity coefficient of LTP, $p_{\max}$: maximum number of LTP pulses (30), $p_1$: LTP pulse number (ranging from 1 to 30).

The LTD data is fitted using:

$$G(p_2) = -\frac{g_{\max} - g_{\min}}{1 - e^{-1/A_d}} \left(1 - e^{\left(\frac{p_2/p_{\max,1} - 1}{A_d}\right)}\right) + g_{\max}, \quad (2)$$

where $A_d$: non-linearity coefficient of LTD, $p_{\max,2}$: maximum number of LTD pulses (30), $p_2$: LTD pulse number (ranging from 1 to 30, obtained after subtracting 30 from the pulse number shown in the x-axis label of Figure 4(b)). $A_p$ ad $A_d$ can be mapped to non-linearity factors $\alpha_p$ and $\alpha_d$ through a mapping table provided with NeuroSim simulator [31,32]. Following the above method of fitting for the experimentally obtained LTP-LTD data of Figure 4(b), we obtain: $A_p = 0.3499$, $\alpha_p = 3.3725$, $A_d = 0.1413$, $\alpha_d = 5.6910$.

Figure 4(c) shows the heatmap of the conductance of the device for all 8 cycles with the colour variation representing how conductance changes with pulse number for each cycle: increasing conductance for LTP, decreasing conductance for LTD. Change in conductance is more sudden for LTP than LTD as also observed in Figure 4(a), (b). This explains lower non-linearity factor reported for LTP compared to LTD. For programming the designed crossbar array from these devices later in the article, we hence use the LTP characteristic and not LTD.

Figure 4(d) shows the retention data over time (stability) of 15 states, selected from the 30 states reported in Figure 4(a), (b), (c). For each state, the conductance value remains stable over $10^3$ s without significant change which confirms the non-volatile nature of these intermediate states between HRS and LRS and the applicability of the device as a non-volatile synapse in a crossbar array. Just like in the case of non-volatile HRS and LRS, here also no prior electro-forming step is needed.

In Section S3, Figure S4, and Table ST4 of Supporting Information, we report how LTP-LTD characteristics of the device can be modified by varying the programming pulse magnitude and duration.

### 3.4 Neural Network Simulations Using LTP-LTD Characteristics of the Device

Having established that our fabricated NiOx device exhibits multiple non-volatile synaptic states (LTP, LTD), we next estimate the classification accuracy of MLPs and CNNs based on crossbar arrays of such devices in on-chip inference mode. Circuit-level design, simulation, and experimental demonstration of MLPs and CNNs using crossbar arrays of NVM synapse devices has been reported extensively in [2,55–61]. Here, we have modeled the MLP and the CNN only through coding in a high-level programming language (Python) while incorporating experimental device characteristics in the synaptic weights of the MLP and the CNN. For this, we first design a MLP or a CNN for a particular data set and train it on the training component of the data set on a GPU (using Python) with all synaptic weights having 32-bit floating point precision. Then, to simulate the on-chip inference of crossbar arrays, the final weights of the trained MLP/ CNN are mapped to weights that can be actually stored in realistic NVM devices with limited fixed-point precision (number of bits stored) and non-zero non-linearity factor of LTP ($\alpha_P$ ). After the mapping, classification accuracy of the MLP/ CNN with the modified weights is obtained on the testing component of the same data set using Python. In our work, this is the reported accuracy corresponding to simulation of on-chip inference using crossbar arrays of synaptic devices.

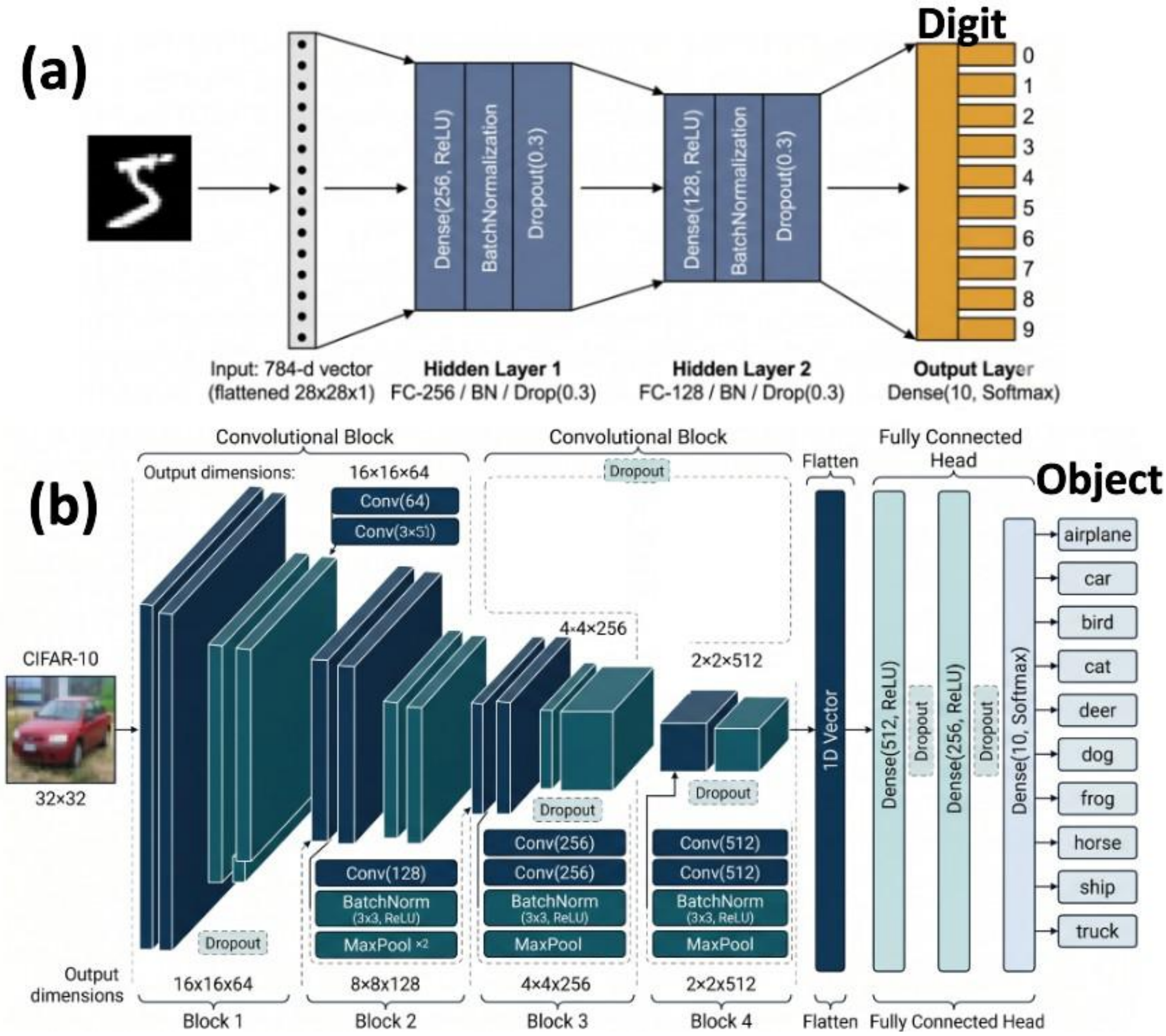

*Figure 5: (a) Schematic of the designed MLP for digit classification using the MNIST data set. (b) Schematic of the designed CNN for object classification using the CIFAR-10 data set.*

Design of the MLP for classification of images of handwritten digits (MNIST data set) is shown in Figure 5(a). There are two hidden layers with each hidden layer containing a dense or fully connected sub-layer, rectified linear units (ReLU) as non-linear activation functions, a batch normalization function, and a drop-out function. More details on these functions can be found in dedicated texts on machine learning and deep learning [28,62,63]. The designed MLP is first trained on 60,000 training samples of the MNIST data set belonging to all ten output classes (ten digits), and the obtained train accuracy is 99.16 %. The obtained on-chip inference accuracy on 10,000 test samples for this MLP, following the weight mapping discussed above, is plotted as a function of $\alpha_P$ and number of bits, which are stored in the NVM synapse, in the form of a heat map in Figure 6(a).

We next consider the CIFAR-10 data set comprising images of different animals and transport vehicles. Given that the classification task here is much more involved than that for MNIST digits, we design a CNN instead of a MLP. Our designed CNN is shown in Figure 5(b). It contains four repetitions of a block that has the following sequence of operations: convolution, batch normalization, convolution, batch normalization, max pooling, drop-out [28,62,63]. The number of extracted features and hence the number of convolutional weight kernels and size of each kernel varies from one block to another as shown in in Figure 5(b). The output of the fourth such block is flattened and then passed into a MLP resembling that in Figure 5(a) with two blocks of dense sub-layers and drop-out functions. The designed CNN is first trained on 50,000 training samples of the CIFAR-10 data set belonging to all ten output classes, and the obtained train accuracy is 92.23 %. The obtained on-chip inference accuracy number on 10,000 test samples for this CNN, following the weight mapping discussed above, is plotted as a function of $\alpha_P$ and number of bits stored in the NVM synapse in the form of a heat map in Figure 6(b).

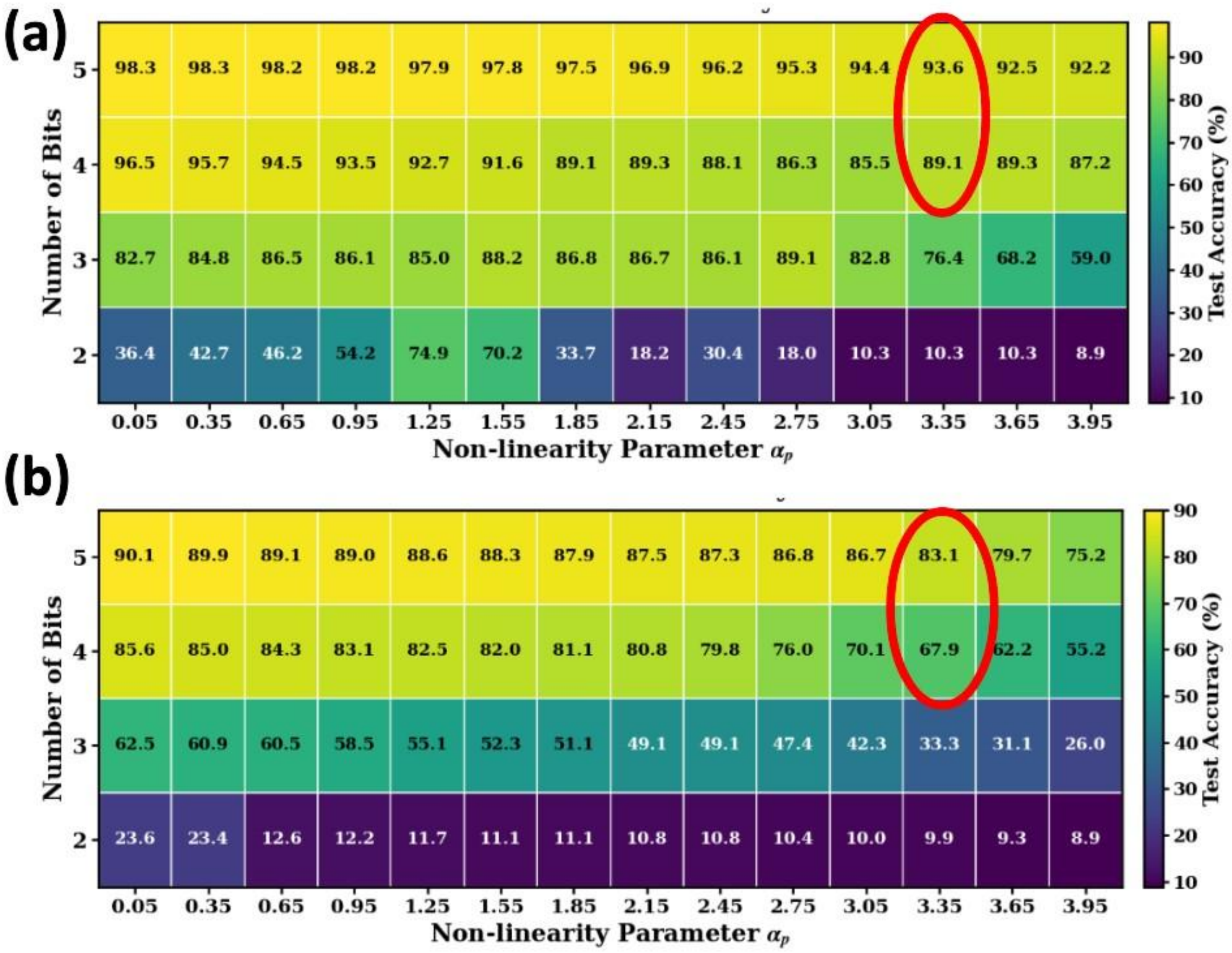

*Figure 6: (a) Heat map showing on-chip inference accuracy of the MLP in Figure 5(a) on MNIST test data set as a function of $\alpha_P$ and number of bits, which are stored in the NVM synapse. (b) Heat map showing on-chip inference accuracy of the CNN in Figure 5(b) on CIFAR-10 test data set as a function of $\alpha_P$ and number of bits in the NVM synapse.*

Experimental data presented in Figure 4(b), (d) show that the NiOx RRAM device in this work can store 4-5 bits of synaptic weight (16-32 intermediate states). Figure 4(b) also shows that non-linearity factor of LTP ($\alpha_P$ ) is approximately 3.35. On-chip inference accuracy numbers for a 4-bit device with $\alpha_P = 3.35$ and a 5-bit device with $\alpha_P = 3.35$ are circled and highlighted in Figure 6(a), (b). We observe for this kind of non-linearity, while a 4-bit device is sufficient to obtain high test accuracy for MNIST digits (since it's a simpler classification task), at least a 5-bit device is needed to obtain reasonably high-test accuracy (above 80 %) for the CIFAR-10 data set.

## 4 Conclusion

Thus, in this article, we experimentally report in solution-processed Ti/Pt/NiO/Ag RRAM devices, the occurrence of non-volatile states without the need for an electro-forming step. Not only we demonstrate forming-free binary switching between stable non-volatile LRS and HRS states in these devices with low cycle-to-cycle and device-to-device variation but also demonstrate forming-free pulse-driven LTP and LTD spanning 30 stable intermediate states (≈5-bit synaptic weight resolution). We use an identical pulse scheme for LTP-LTD. Elimination of the forming step and use of this identical pulse scheme make peripheral circuit design for programming the synapses in the crossbar array simple. As evidenced by Table 1 and Table 2, to the best of our knowledge, this is one of the few reports of forming-free occurrence of non-volatile binary states (HRS, LRS) and non-volatile intermediate states in NiO RRAM devices, including both devices where NiO layer is deposited by sputtering and where NiO layer is deposited by solution processing [9, 10, 10, 15, 16, 16, 18, 18].

**Acknowledgment**

Debanjan Bhowmik and Merlyne De Souza acknowledge financial support from Ministry of Education India (MoE) through the Scheme for Promotion of Academic and Research Collaboration (SPARC) program (project code: SPARC/2025- 2026/P4357). Debanjan Bhowmik also acknowledges financial support from Ministry of Education India (MoE) through the MoE-STARS program (project code: MoE-STARS/STARS-2/2023-0257) and from Anusandhan National Research Foundation (ANRF) through the ANRFMATRICS program (project code: MTR/2023/000851). The authors also thank Bhaswar Chakrabarti, Indian Institute of Technology Madras, for discussions.

**Supporting Information**

Supporting Information accompanying this article can be found below.

**Supporting Information for “A Solution-Processed Ti/Pt/$NiO_x$/Ag RRAM Device with Multiple Forming-Free Non-Volatile Intermediate States”**

Abhrajit Roy[1], Chopali Chanchal Sahu[1], Rudraksh Sangore[1], Raman Hissariya[1], Poulomi Chakrabarty[2], Harsh Roniyar[1], Merlyne De Souza[3], and Debanjan Bhowmik[1,4]

[1]*Department of Electrical Engineering, Indian Institute of Technology Bombay, Mumbai, Maharashtra 400076, India*

[2]*Department of Computer Science and Engineering, Shiv Nadar University Chennai, Kalavakkam, Tamil Nadu 603110, India*

[3]*School of Electrical and Electronic Engineering, The University of Sheffield, Western Bank, Sheffield, S10 2TN, United Kingdom*

[4]*Centre for Semiconductor Technologies (SemiX), Indian Institute of Technology Bombay, Mumbai, Maharashtra 400076, India*

**Section S1: Material Deposition and Characterization Methods, Device Fabrication and Characterization Methods**

The steps we have followed to deposit the thin film stack and fabricate the RRAM devices for our work have been discussed in detail in the main text. Here, in Fig. S1, we show a schematic illustrating preparation of the NiO precursor solution gel for solution-processingbased deposition of the materials stack. In Fig. S2, we show the overall fabrication procedure of the Ti-Pt/NiO/Ag device. The main steps in the fabrication are all shown in this schematic: sputtering of bottom electrode (BE) (Ti/Pt), subsequent spin coating of NiO precursor solution (as prepared in the schematic of Fig. S1), and then deposition of top electrode (TE) (Ag) through optical-lithography-based patterning and sputtering.

Below we have provided some additional information on the aforementioned material deposition and device fabrication steps, which we have not provided in the main text. This includes instrument details. We also provide additional details on our material characterization and device characterization techniques (including instrument details), which we have not provided in the main text.

Details of the solution process we followed have been provided in the main text. For deposition of the bottom electrode (Ti/Pt) and top electrode (Ag), DC sputtering has been carried out using a high vacuum magnetron sputtering tool from AJA International (Orion series). For deposition of the top electrode, patterning has been carried out through photolithography using the optical lithography system (MicroWriter ML3. For this photolithography, lift-off technique using a positive photoresist: Microposit S1813.

Details of the top electrode and bottom electrode in our device stack are further listed in Table ST1 and compared with that used in other related reports on solution-processed NiO devices in the same table. Details of our solution processing and spin coating steps to deposit the oxide layer are provided in Table ST2 and compared with that followed in related reports on solution-processed NiO devices in the same table. Details of the materials stack and fabrication process for sputtered NiO devices is provided in Table ST3.

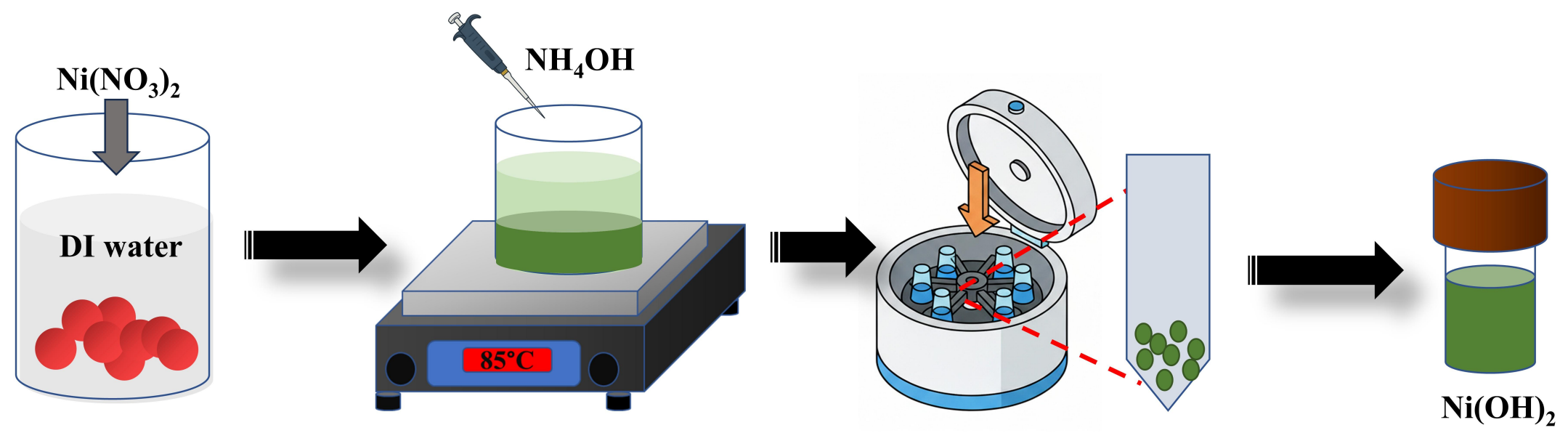

***Figure S1:*** *Schematic illustrating preparation of the NiO precursor solution gel ($Ni(OH)_2$) for solution-processing-based deposition of the materials stack*

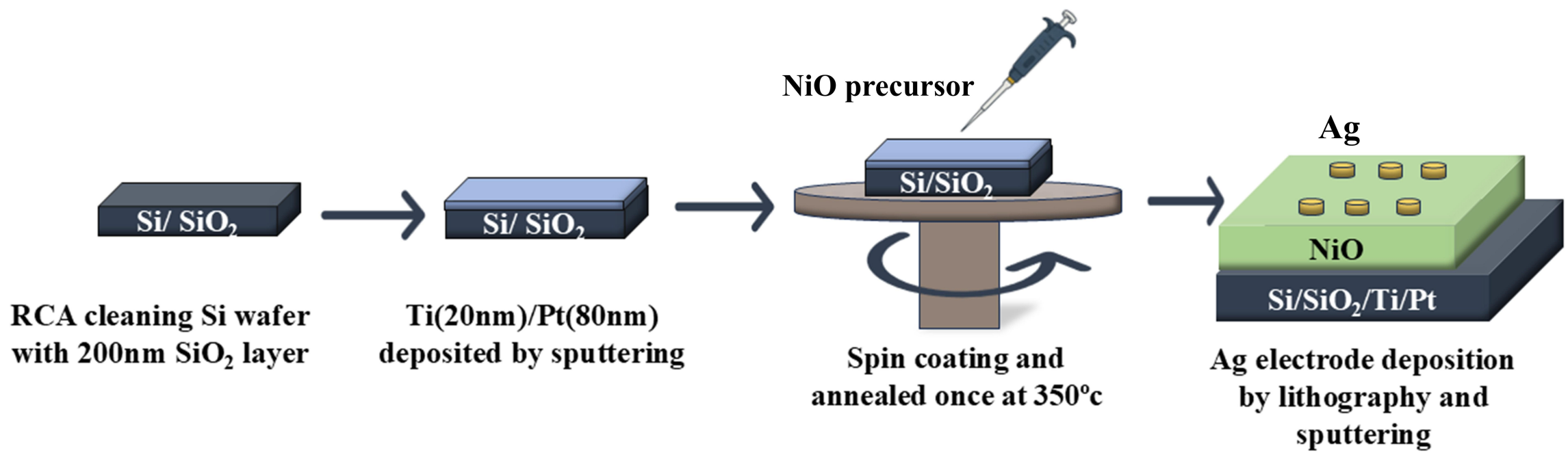

***Figure S2:*** *Schematic showing overall procedure for device fabrication: sputtering of bottom electrode (Ti/Pt) followed by spin coating of NiO precursor solution (as prepared in Fig. S1) and then deposition of top electrode (Ag) through optical-lithography-based patterning and sputtering*

To characterize the thin film stack, high resolution X-ray diffraction (HRXRD) has been carried out using a HRXRD tool from Rigaku SmartLab. A continuous scan mode has been used with the scan rate being 2°/min and step increment being 0.02°.

Next, X-ray reflectivity (XRR) measurements have been performed on the same thin film stack using the same XRD tool in a continuous scan configuration over a $2\theta$ range of 0.3° to 5.0°. The scan rate for XRR is chosen to be 0.05°/min with a step increment of 0.005°.

Furthermore, X-ray photoelectron spectroscopy (XPS) has been performed using a PHI 5000 VersaProbe II system equipped with a monochromatic Al K$\alpha$ source ($h\nu$ = 1486.7 eV). The measurements have been carried out with an X-ray beam size of 100 $\mu$m, operating at 25 W and 15 kV. A pass energy of 58.7 eV and an energy resolution of 0.125 eV have been used with the hemispherical analyzer to obtain high-quality spectra.

Results from our XRD, XRR, and XPS characterization of the thin film stack have been presented in the main text including Fig. 1 of the main text.

Electrical characterization of the NiO RRAM device has been carried out using the Keysight B1500A semiconductor parameter analyzer. All measurements have been carried out at room temperature under ambient conditions in a probe station. Voltage pulses have been applied on the device using the waveform generator/ fast measurement unit (WGFMU) associated with the semiconductor parameter analyzer.

| **Reference** | **Substrate** | **Bottom electrode (BE), thickness** | **BE deposition technique** | **Top electrode (TE), thickness** | **TE deposition technique** |
|---|---|---|---|---|---|
| Li *et al.* [1] | $Si/SiO_2$ | Ti (50 nm) / Pt (20 nm) | E-beam evaporation | Ag (160 nm) | E-beam evaporation |
| Chu *et al.* [2] | $Si/SiO_2$ | Ti (50 nm) / Pt (20 nm) | E-beam evaporation | Ag (thickness not reported) | E-beam evaporation |
| Hadiyal *et al.* [3] | glass | Au (thickness not reported) | Not mentioned | Au (thickness not reported) | Not mentioned |
| Kanan *et al.* [4] | ITO | ITO | substrate used as BE | Al (120 nm) | e-beam evaporation |
| Ahn *et al.* [5] | Si | Nb nanopin | wet etching | Nb (100 $\mu$m, 50 nm) | sputtering |
| Park *et al.* [6] | glass | ITO | Not mentioned | Ag | thermal evaporator |
| **This work** | $Si/SiO_2$ | Ti (20 nm) / Pt (80 nm) | DC sputtering | Ag (100 nm) | DC sputtering |

*Table ST1: For solution-processed NiO RRAM devices reported in literature and that reported in our work, a comparison of their top electrode (TE) and bottom electrode (BE) deposition techniques is carried out here*

**Section S2: Cycle-to-Cycle Variations in Digital Switching Characteristics**

On the Ti/Pt/NiO/Ag RRAM device fabricated for this work (top electrode diameter: 50 $\mu$m), we have repeated the I-V characteristic across 1000 cycles and studied its digital switching between high resistance state (HRS) and low resistance state (LRS). I-V sweeps corresponding to the 1st, 51st, 101st, 151st, 201st, 251st, 301st, 351st, and 401st cycle are plotted in Fig. S3.

**Section S3: Dependence of LTP-LTD on Pulse Magnitude and Duration**

In Fig. 3 of the main text, we have reported the LTP-LTD characteristics of the device for +/- 0.3 V pulses of 200 ns duration. Sticking to this identical pulse scheme, here we vary the magnitude and duration of the pulses and plot in Fig. S4(a) how LTP-LTD changes. Fig. S4(b) shows fitted LTP-LTD characteristics, following equation 1 and 2 (of the main text) with appropriate non-linearity factors $\alpha_P$ and $\alpha_D$, on the experimental data of Fig. S4(a). The obtained non-linearity fac tors are listed in the legend of Fig. S4(b) and again listed in Table ST4.

| Ref. | Oxide layer (thickness) | Preparing NiO precursor solution | $O_2$ plasma before precursor deposition on first | NiO precursor deposition method | Baking temperature and time | Number of repetitions of precursor depositio | $O_2$ plasma at every precursor deposition step? | Final annealing temperature and time |
|---|---|---|---|---|---|---|---|---|
| | | | **time?** | | | **n** | | |
| Li *et al.* [1] | NiO (50 nm) | $Ni(NO_3)_2$ + 2methoxyethanol + $NH_4OH$ + acetylacetone | Yes | Spin coat | 300°C, 30 min | 5 | Yes | 350°C, air, 30 min |
| Chu *et al.* [2] | NiO (50 nm) | $Ni(NO_3)_2$ + 2methoxyethanol + $NH_4OH$ + acetylacetone | Yes | Spin coat, 4300 rpm | 200°C, 30 min | 5 | Not mentioned | 200°C, air, 1 h |
| Hadiyal *et al.* [3] | NiO (thickness not reported) | $Ni(NO_3)_2$ + DI water + NaOH | No | Drop cast | 150°C, ~2–4 h | several | Not mentioned | 200°C, 10 h |
| Kanan *et al.* [4] | NiO (70 nm) | $C_6H_4NiO_4 \cdot 4H_2$ + 2-methoxyethanol + 2-aminoethanol | Not mentioned | Spin coat, 3000 rpm, 30 s | 120°C, 5 min | 5 | Not mentioned | 475°C, 2 h |
| Ahn *et al.* [5] | NiO 50 nm | $Ni_2(CO_3)(OH)_2$ solution | Not mentioned | Spin coat | Not mentioned | Not mentioned | Not mentioned | 400°C, 1 h |
| Park *et al.* [6] | NiO ~23 nm | $Ni(NO_3)_2 \cdot 6H_2O$ + ethylene glycol + ethylenediamine | Not mentioned | 3000 rpm, 60 s | 150°C, 5 min | Not mentioned | Not mentioned | 300°C, 60 min |
| **This work** | NiO (17 nm) | $Ni(NO_3)_2$ + DI water + $NH_4OH$ | Yes | Spin coat, 2500 rpm | 120°C, 10 min | 5 | Yes | 350°C, 1 h |

***Table ST2:*** *For solution-processed NiO RRAM devices reported in literature and that reported in our work, a comparison of experimental details regarding solution-processing-based deposition of the oxide layer in the devices is carried out here.*

| **Device / Reference** | **Substrate** | **Bottom electrode (BE), thickness** | **BE deposition technique** | **Oxide layer, thickness** | **Sputtering target for oxide layer** | **Nature of sputtering for oxide layer** | **Final annealing temp. and time** | **Top electrode (TE), thickness** | **TE deposition technique** |
|---|---|---|---|---|---|---|---|---|---|
| Sarkar *et al.* [7] | $Si/SiO_2$ | Pt | commercial | $NiO_y$ (20 nm) / $NiO_x$ (20 nm) | 99.99 % pure NiO target | RF sputtering | – | Cu (100 nm) | Just mentioned shadow masking |
| Ma *et al.* (Ref. 10) | Si | Pt | Not mentioned | NiO (50 nm) | nickel target (99.99 %) | DC magnetron sputtering | 550 ˚C for 10 min after deposition | Cu (area 100 × 100 $\mu$m) | magnetron sputtering |
| Mondal *et al.* [8] | Si | Ti (10 nm) / Pt (70 nm) | Electron beam evaporation | $NiO_x$ (96 nm) | pure Ni target | sputtering technique | 350 ˚C, 400 ˚C, and 450 ˚C, respectively for 1 hr in air | Pd (200 nm) | sputtering |
| Hsin *et al.* [9] | oxidized silicon (100) substrate | W (3 nm) / Pt (4 nm) / Co | DC sputtering | NiO (30 nm) | NiO target | RF sputtered | – | no separate metal top | – |
| | s | (0.8 nm) | | | | | | electrod e | |

***Table ST3:*** *Comparison of stack and deposition details for sputtered NiO RRAM devices reported in literature.*

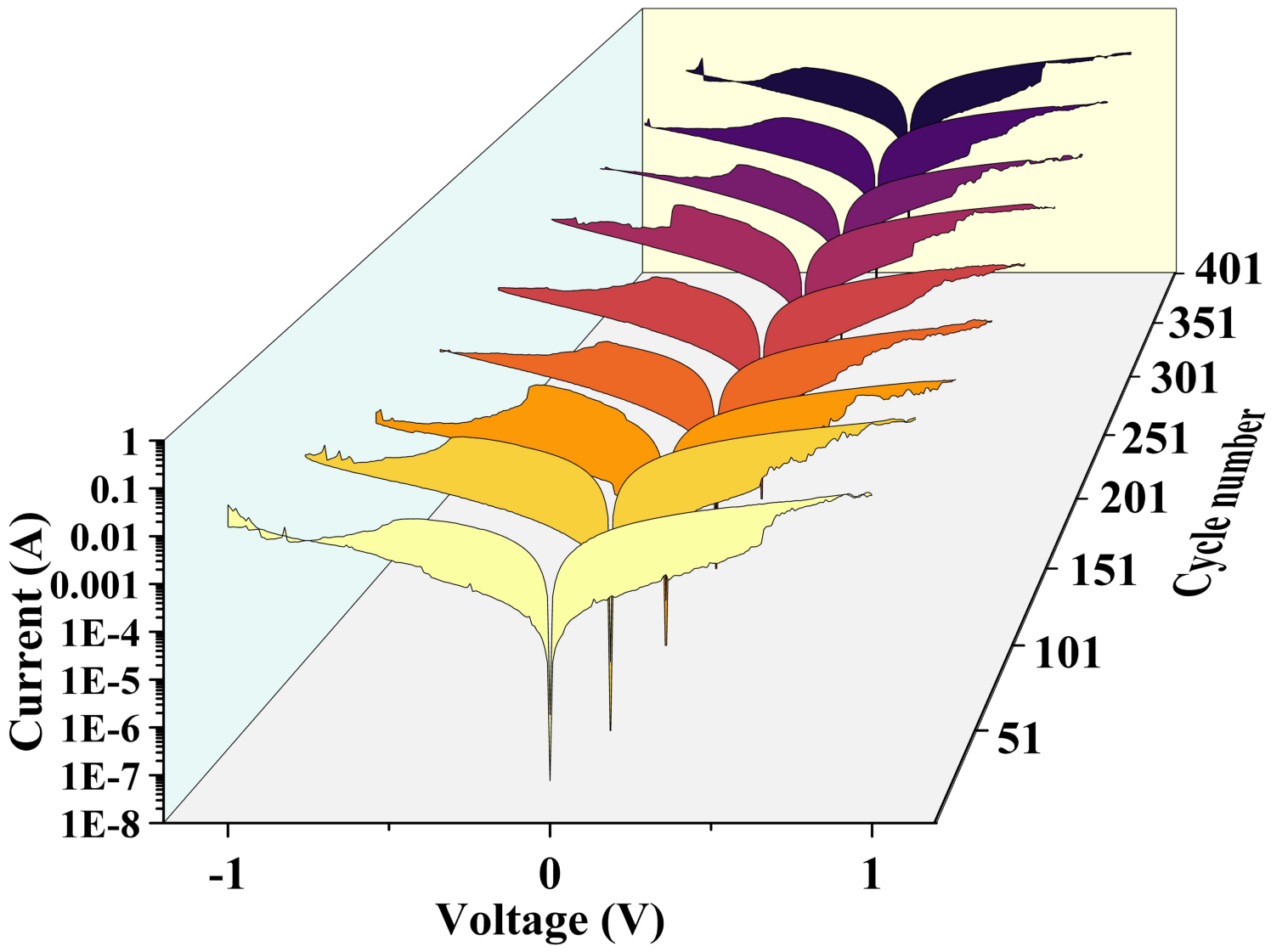

***Figure S3:*** *I-V characteristics of the Ti-Pt/NiO/Ag RRAM device (top electrode diameter: 50 μm) are plotted here corresponding to the 1st, 51st, 101st, 151st, 201st, 251st, 301st, 351st, and 401st cycle.*

| **Pulse Condition** | **LTP ($A_p$)** | **LTP ($\alpha_p$)** | **LTP ($A_d$)** | **LTD ($\alpha_d$)** |
|---|---|---|---|---|
| ±0.3 V, 500 ns | 0.2771 | 3.9305 | 0.1990 | 4.7812 |
| ±0.4 V, 500 ns | 0.3683 | 3.2548 | 0.3098 | 3.6583 |

***Table ST4:*** *Summary of the pulse conditions and corresponding LTP-LTD fitting parameters obtained from Fig. S4 (a), (b).*

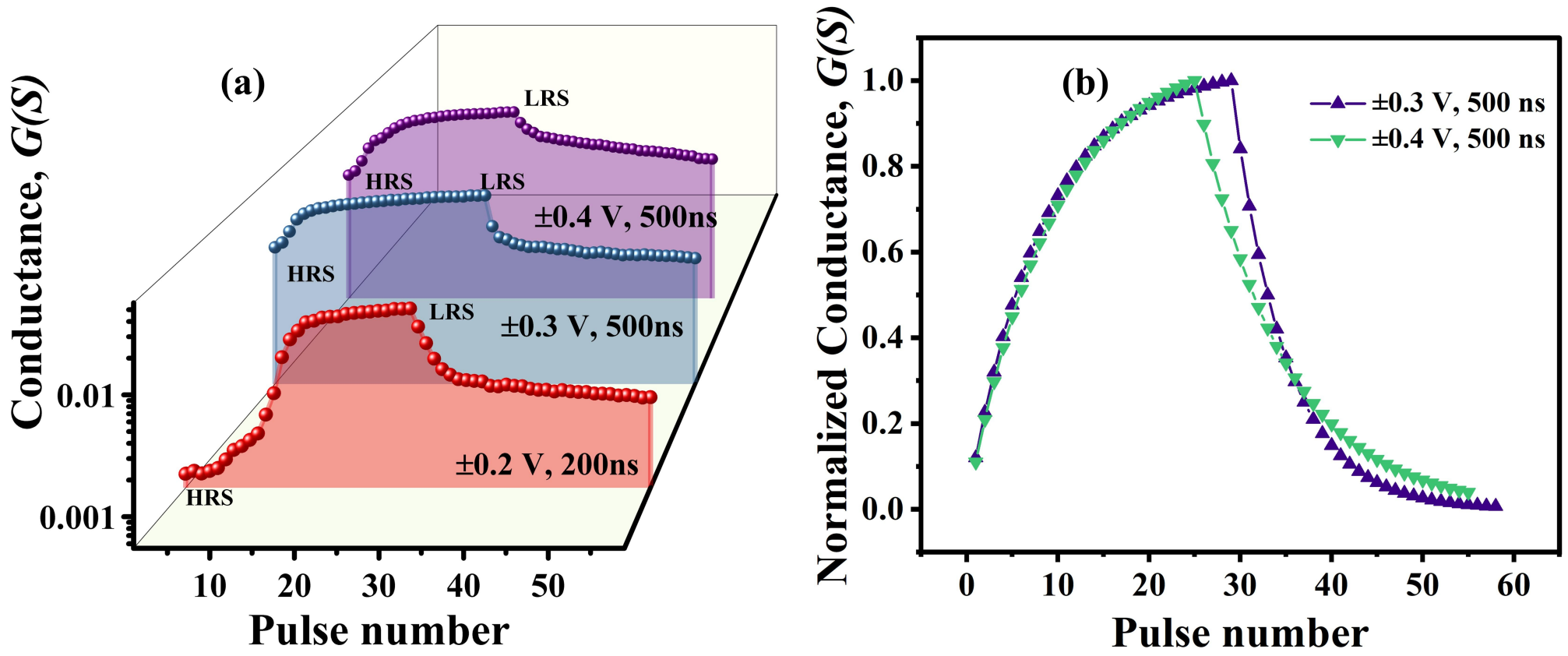

***Fig. S4:*** *(a) LTP-LTD charcteristics of the NiO-based RRAM device under identical positive pulses for LTP and identical negative pulses for LTD where the pulse magnitude and duration is varied (b) Fitting of equations 1 and 2 of the main text on the LTP-LTD characteristics of (a), non-linearity factors are listed in the legend.*